\documentclass[pra,twocolumn,showpacs,superscriptaddress,floatfix,nofootinbib]{revtex4-2}

\usepackage{color}
\usepackage{hyperref}  
\usepackage{graphicx}
\usepackage{amsfonts}
\usepackage{footmisc}

\usepackage{fancyhdr}
\let\a=\alpha \let\b=\beta  \let\g=\gamma  \let\d=\delta 
  \let\h=\eta   \let\th=\theta  \let\l=\lambda
\let\m=\mu    \let\n=\nu    \let\x=\xi     \let\p=\pi    \let\r=\rho
\let\s=\sigma    \let\f=\varphi 

 \let\D=\Delta  \let\L=\Lambda

\font\tenmib=cmmib10\font\sevenmib=cmmib7\font\fivemib=cmmib5%
\mathchardef\Bl   = "0515  

\def\Bl   {{\mbox{\boldmath$ \lambda$}}}

\def\Bff  {{\mbox{\boldmath$ \varphi$}}}

\def\BDpr {{\mbox{\boldmath$ \partial$}}}

\def\eqalign#1{\null\,\vcenter{\openup\jot
  \ialign{\strut\hfil$\displaystyle{##}$&$\displaystyle{{}##}$\hfil
      \crcr#1\crcr}}\,}

\def\AA{{\mathcal A}}\def\CC{{\mathcal C}}\def\FF{{\mathcal F}}
\def\EE{{\mathcal E}}\def\DD{{\mathcal D}}\def\TT{{\mathcal T}}
\def\NN{{\mathcal N}}
\def\LL{{\mathcal L}}\def\HH{{\mathcal H}}
\def\RR{\cal R}

\def\uu{{\V u}}\def\kk{{\V k}}\def\xx{{\V x}}\def\ff{{\V f}}

\def\qq{{\V q}}\def\pp{{\V p}}
\def\T#1{{#1_{\kern-3pt\lower7pt\hbox{$\widetilde{}$}}\kern3pt}}

\def\ie{{\it i.e.\ }}
\def\dpr{{\partial}}
\def\defi{{\buildrel def\over=}}

\def\otto{\,{\kern-1.truept\leftarrow\kern-5.truept\to\kern-1.truept}\,}
\def\Pprod{\prod^{\kern-1mm\raise.0mm\hbox{$\leftarrow$}}}
  
\newdimen\xshift \newdimen\xwidth \newdimen\yshift \newdimen\ywidth

\def\ins#1#2#3{\vbox to0pt{\kern-#2pt\hbox{\kern#1pt #3}\vss}\nointerlineskip}

\def\eqfig#1#2#3#4#5{
\par\xwidth=#1pt \xshift=\hsize \advance\xshift
by-\xwidth \divide\xshift by 2
\yshift=#2pt \divide\yshift by 2
{\hglue\xshift \vbox to #2pt{\vfil
#3 \includegraphics{#4.eps}
}\hfill\raise\yshift\hbox{#5}}}

\def\V#1{{\bf #1}}
\def\lis#1{{\overline#1}}
\let\wt=\widetilde

\def\tende#1{\,\vtop{\ialign{##\crcr\rightarrowfill\crcr
 \noalign{\kern-1pt\nointerlineskip} \hskip3.pt${\scriptstyle
   #1}$\hskip3.pt\crcr}}\,}
\def\eg{{\it e.g.\ }}

\def\0{\noindent}
\def\*{\vskip2mm}
\def\media#1{\langle #1 \rangle}
\def\Eq#1{\label{#1}}
\def\equ#1{(\ref{#1})}

\font\titolo=cmbx12%
\def\iniz{\setcounter{equation}{0}}
\def\be{\begin{equation}}\def\ee{\end{equation}}
\renewcommand{\theequation}{\arabic{section}.\arabic{equation}}

\newcounter{appendice}
\def\APPENDICE#1{
\setcounter{appendice}{#1}
\appendix
\renewcommand{\theequation}{\Alph{appendice}.\arabic{equation}}%
\renewcommand{\thesection}{\Alph{appendice}}%
}

\def\alert#1{{\color{ired}#1}}
\definecolor{iblue}{RGB}{65,105,225}
\definecolor{ired}{RGB}{220,20,60}
\definecolor{igreen}{RGB}{50,205,50}
\definecolor{ipurple}{RGB}{75,0,130}
\definecolor{iochre}{RGB}{218,165,32}
\definecolor{iteal}{RGB}{51,204,204} 
\definecolor{imauve}{RGB}{204,51,153}

\def\pagina{\vfill\eject}

\def\ap{{\it a priori }}
\iniz

\begin{document}

\let\titolo=\bf

\alert{\centerline{\bf  Viscosity, Reversibillity,
    Chaotic Hypothesis,  }}
  \alert{\centerline{\bf   
    Fluctuation Theorem and Lyapunov Pairing
      }}
  \vskip1mm
 
\centerline{\bf Giovanni Gallavotti} \centerline{\today}

{\vskip3mm}
\noindent {\bf Abstract}: {\it Incompressible fluid equations are
  studied with UV cut-off and in periodic boundary
  conditions. Properties of the resulting ODEs holding uniformly
  in the cut-off are considered and, in particular, are
  conjectured to be equivalent to properties of other time
  reversible equations.  Reversible equations with the same
  regularization and describing equivalently the fluid, and the
  fluctuations of large classes of observables, are examined in
  the context of the ``Chaotic Hypothesis'', ``Axiom C'' and the
  ``Fluctuation Theorem''.}  {\vskip3mm}

\def\SEC{On the equations}
\section{\SEC}
\label{sec1}
\def\Dot#1{{\bf\dot#1}}

The incompressible Euler equation, denoted $\bf E$, in a periodic
container $\TT^d=[0,2\p]^d,\,d=2,3$, for a smooth velocity field
$\uu(x), x\in \TT^d$ is:

\be\kern-3mm\eqalign{
  &\Dot\uu(x)=\kern-1mm-(\T \uu(x)\cdot
  \T\BDpr_{\raise 2mm \hbox{$x$}})
  \uu(x)\kern-1mm-\BDpr_x P(x),\cr
  &\BDpr_x\cdot\uu(x)=0,\quad \int_{\TT^d} dx\,
  \uu(x)=\V 0\cr}\Eq{e1.1}\ee
where $P=-\sum_{i,j=1}^d\D^{-1}(\dpr_i u_j\, \dpr_j u_i)$
is the 'pressure' and  $\D$=Laplace operator.

It is also useful to consider the $\bf E$ equations from the
``Lagrangian viewpoint'': a configuration of the fluid is
described by assigning the dispacement $x=q_\x$ of a fluid
element, from the reference position $\x\in\TT^d$, and the
velocity $\dot q_\x$ of the same fluid element. So the state of
the fluid is $(\qq,\Dot q)$ where $\qq$ is a smooth map of
$\TT^d$ to itself and $\Dot q$ is a smooth vector field on
$\TT^d$ with $\int \Dot q_\x d\x=0$. Denote $\FF$ the space of
the dynamical configurations $(\qq,\Dot q)\in Dif(\TT^d)\times
Lin(\TT^d)=\FF$ where $Dif(\TT^d)$ is the set of diffeomorphisms
of $\TT^d$ and $Lin(\TT^d)$ the space of the vector fields on
$\TT^d$.

Actually we concentrate on the subspace of $(\qq,\Dot q)\in
(SDif(\TT^d)\times SLin(\TT^d))\defi S\FF\subset\FF$ where the
evolution of an {\it incompressible} fluid takes place:
$SDif(\TT^d)$ being the {\it volume preserving} diffeomorphisms
and $SLin(\TT^d)$ the {\it $0$-divergence} vector fields.

A $(\qq,\Dot\qq)=\{q_\x,\dot q_\x\}_{\x\in\TT^d}\in\FF$ should be
regarded as a set of Lagrangian coordinates labeled by
$\x\in\TT^d$. And the equations 
Eq.\equ{e1.1} can be derived from a Hamiltonian in canonical
coordinates $(\qq,\pp)\in \FF$ which is {\it quadratic} in $\pp$
and which generates motions in $\FF$ evolving leaving $S\FF$
invariant.  Therefore the motion in $\FF$ is a ``geodesic
motion'' (\ie a motion generated by a Hamiltonian quadratic in
the momenta).

A key remark is that the motions that follow initial data in
$S\FF$ remain, {\it as long as the evolution is defined and
  smooth},\footnote{\small This might be a very short time.} in
$S\FF$, \cite{Ar966b,TT010}, \ie $S\FF$ is an invariant surface
in $\FF$. And the equations of motion that $H$ generates can be
written (using incompressibility of $(\qq,\pp)\in S\FF$) as:
\be\eqalign{ \dot p_\x=-\BDpr_{\scriptstyle q_\x}\kern-1mm
  Q(\qq,\pp)_\x,\quad \dot q_\x=p_\x \cr} \Eq{e1.2}\ee
where $Q_\x$ is $\qq$-dependent and quadratic in $\pp$, see
Appendix \ref{seca}. 

Since $\dot p_\x=\dpr_t p_{\x}+(\T p_{\raise2mm\hbox{$\x$}}
\cdot\T\dpr p_\x)$, setting $x=q_\x,u(x)=p_\x$ and 
$P(x)=\BDpr_{\scriptstyle q_\x}\kern-1mm Q(\qq,\pp)_\x$, the
equations become:
\be\eqalign{&\dot q_\x=p_\x, \quad
  \dpr_t u(x)+((\T u\cdot\T\dpr) u)(x)
  =-\dpr P(x)\cr}
\Eq{e1.3}\ee
with $\BDpr\cdot\uu=0$, and $P$ as above. The {\it Lagrangian
  form} of Euler's equations, Eq.\equ{e1.2} or
\equ{e1.3}, will be called $\bf E^*$.  See Appendix \ref{seca}.

The above ``geodesic'' formulation of $\bf E,E^*$ will be used to
exhibit symmetry properties of Euler's equation which may be
relevant also for the IN ({\it irreversible Navier-Stokes})
equations:
\be \dpr_t \uu(x)+((\T \uu\cdot\T\dpr) \uu)(x)
=\n \D\uu(x)-\dpr P(x)+\ff(x)\Eq{e1.4}\ee
with the conditions $\BDpr\cdot\uu=0$, $\int_{\TT^d}\uu=0$.

\def\SEC{Ultraviolet regularization}
\section{\SEC}
\label{sec2}\def\eE{{\bf e}}
\iniz

Here we study the regularized version, see below, of $\bf E$ or
IN, Eq.\equ{e1.1},\equ{e1.4}, obtained by requiring that the
Fourier's transform $\uu_\kk$ of $\uu$ does not vanish only for
modes $\kk$ with components $\le N$.

We shall {\it focus on properties
  of the solutions which hold uniformly in the cut-off $N$}: the
space of such $\uu$'s with $0$ divergence ($\BDpr\cdot\uu=0$) and
$0$ average ( $\int_{\TT^d}\uu(x)=0$) will be denoted $\CC_N$.

Therefore the equation in dimension $d=2,3$ is expressed in terms
of complex scalars $u_{\b,\kk}=\lis u_{\b,-\kk},\,
\b=1,\ldots,d,\,\kk\in Z^d,\, |\kk_\b|\le N$: thus the 
number of real coordinates is $\NN=4N(N+1)$ in 2D 
and $\NN=2(4N^3+6N^2+3N)$ in 3D and $\NN$ will be the dimension
of the phase space $\CC_N$.

For instance in 3D choose, for each $\kk\ne0$, two unit vectors
$\eE_\b(\kk)=-\eE_\b(-\kk),\, \b=1,2$, mutually orthogonal and
orthogonal to $\kk$; data are combined to form a velocity
field:
\be
\eqalign{
  &\uu(\xx)=\sum_{0<|\kk|\le N}
  \uu_\kk e^{-i\kk\cdot\xx},\ \ \kk=(k_\b)_{\b=1,2,3}\cr
  & \uu_\kk=\sum_{\b=1,2} i\,u_{\b,\kk} \,\eE_{\b}(\kk),\qquad
  \kk\cdot\eE_\b(\kk)=0\cr
  }\Eq{e2.1}\ee
with $|\kk|=\max_j |k_j|, u_{-\kk,j}=\lis u_{\kk,j}$.

Define $D_{\kk_1,\kk_2,\kk}^{\b_1,\b_2,\b}=-(\V
e_{\b_1}(\kk_1)\cdot \kk_2) (\V e_{\b_2}(\kk_2)\cdot \V
e_{\b}(\kk))$. Introduce also forcing $\ff=\sum_{\kk,\b}
i f_{\kk,\b} \eE_\b(\kk) e^{-i\kk\cdot\xx}$ and viscosity in the
form $-\n \kk^2 \uu_\kk$. 

The IN equations Eq.\equ{e1.4} become, if $\kk^2\defi
k_1^2+k_2^2+k_3^2$ and the sum is restricted to
$|\kk|,|\kk_1|,|\kk_2|\le N$:
\be\kern-2mm \dot u_{\b,\kk}=\kern-2mm
  \sum_{\b_1,\b_2\atop\kk=\kk_1+\kk_2}\kern-2mm
  D_{\kk_1,\kk_2,\kk}^{\b_1,\b_2,\b}\,
  \,u_{\b_1,\kk_1}u_{\b_2,\kk_2}
  -\n\kk^2 u_{\b,\kk}+ f_{\b,\kk}\Eq{e2.2}\ee
which will define the {\it regularized IN equation}.

The 2D case is similar but simpler: no need for the labels $\b$,
and $\eE(\kk)$ can be taken $\frac{\kk^\perp}{||\kk||}$.

The coefficients $D_{\kk_1,\kk_2,\kk_3}^{\b_1,\b_2,\b_2}$ can be
used to check that if $\n=0,\ff=\V0$ then for all
$\uu\in\CC_N$:\footnote{\small The symmetries of $D$ arise from
  the identities $\int ((\T\uu\cdot\T\BDpr)\uu)\cdot\uu=0$ and
  $\int (\uu\cdot\BDpr\T\uu)\cdot(\T\BDpr\wedge\T\uu)=0$, by
  integration by parts.}
\be \frac{d}{dt}\int_{\TT^d}\uu(\xx)^2
d\xx=0,\ \frac{d}{dt}\int_{\TT^d}\kern-3mm\uu(\xx)\cdot(\BDpr
\wedge\uu(\xx))d\xx=0\Eq{e2.3}\ee
As is well known, the first of Eq.\equ{e2.3} leads to the {\it a
  priori}, $N$-independent, bounds for the solutions of the $\bf
E$ and IN equations:
\be\kern-3mm
||\uu^{X,N}(t)||^2_2\le \max(E_0, (\frac{F_0}\n)^2),\ X={\bf E},IN
\Eq{e2.4}\ee
satisfied (for all X) by solutions $t\to\uu^{X,N}(t)\defi
S^{X,N}_t\uu$, in terms of $E_0=
||\uu(0)||_2^2=\sum_{\b,\kk}|u_{\b,\kk}|^2$ and $F_0=||\ff||_2$.

From now on the {\it cut-off N will be kept constant} and the
solution of the equations will be denoted simply $S_t\uu$ dropping
the $X,N$ as superscript of the solution map $S_t$. Only when not
clear from the context a superscript $\bf E$ or IN or a label $N$
will be added to clarify whether reference is
made to the evolution, or to its properties, following $\bf
E$ or IN equation with cut-off $N$.

By scaling, the equation can and will be written in a fully
dimensionless form in which $||\ff||_2=1$.

The Jacobian of the Euler flow $S^{{\bf E},{\rm N}}_t\uu$ with UV
cut-off $N$ is more easily written, without using the Fourier's
transform representation of $\uu$, directly from Eq.\equ{e1.1}
and, see Appendix \ref{secb}, is the {\it sum} of the following
convolution operator on $(\f_j(x))_{j=1}^d= \Bff\in
L_2(\TT^d)\times R^d$:
\be \frac{\dpr \dot u_i(x)}{\dpr u_{j}(y) }
=-{\cal P}\d(x-y) \dpr_{y_j}u_{i}(y),\ i,j=1,..,d,
 \Eq{e2.5}\ee
{\it plus} an antisymmetric operator on the same space;
 here ${\cal P}$ is the orthogonal projection, in the
$L_2(T^d)\times R^d $ metric, on the divergenceless fields $\Bff$.
\footnote{\small Other projections could be used: this is
  convenient to follow the analysis in \cite{Ru982}.}

The operator acts on the fields $\Bff$ with $0$ divergence (this
is used in deriving Eq.\equ{e2.5} to discard contributions
that vanish on the divergenceless fields $\Bff$): and in the end
its {\it symmetric} part is $\cal P$ times the multiplication
operator, on $0$-divergence fields $(\f_j(x))_{j=1}^d=
\Bff\in L_2(\TT^d)\times R^d$, by:
\be W_{i,j}(x)=\frac12\Big(\dpr_{x_j}u_{i}(x)
+\dpr_{x_i}u_{j}(x)\Big)\Eq{e2.6}\ee
\ie $\cal P$ times the operator $(J\Bff)_i(x)=\sum_j
W_{i,j}(x)\f_j(x)$.  \*
 
Introducing also viscosity (and forcing, which however does not
contribute) Eq.\equ{e2.6} immediately leads to express the
symmetric part of the Jacobian of the regularized IN,
irreversible Navier-Stokes, as $\cal P$ times $J_{i,j}^\n=\n
\d_{i,j}\D +W_{i,j}$.

Defining, for $d=2,3$, $w(x)^2=\frac{d-1}{4d} \sum_{i,j=1}^d
W_{i,j}(x)^2$ the inequality $J^\n\le \n\D+w(x)$, derived in
\cite{Ru982,Li984} for the nonregularized IN equation, remains
valid for the regularized one and leads to the bound,
\cite{Ru982,Li984}: \*

\0{\it Theorem:} {\it the sum of the time averages of the first $p$
  eigenvalues of the (Schr\"odinger operator) $\n D+w(x)$ yields
  an upper bound to the sum of the first $p$ Lyapunov exponents
  (of any invariant distribution on $\FF$) of the flow $S_t$}.
\*

Remark: Lyapunov exponents depend on the invariant distribution
used to select data: here they will be defined as the time
averages of the eigenvalues of the symmetric part of the Jacobian
of the evolution equation, \cite{Ru982}.\label{lyapunov} The
$u$-dependent non averaged eigenvalues will be called {\it local
Lyapunov exponents}.

\def\SEC{Reversible
  equations}
\section{\SEC}
\label{sec3}
\iniz

The theory of nonequilibrium fluctuations has led to studying
phenomena via equations considered equivalent (at least for some
of the purposes of interest) to the ``fundamental'' ones. 

Thus new non-Newtonian forces have been added to systems of
particles claiming that the values of important quantities would
have the same values as those implied by the fundamental
equations, even in cases in which the modification was drastic:
with the advantage, in several cases, of greatly facilitating
simulations, \cite{No984,EM990,Ho999}.\footnote{By fundamental I
mean here equations based on Newton's principles as theoretical
base for studying quantities like transport coefficients \eg in a
shear flow model, \cite{ECM990,ECM993}, or properties of heat
conduction, \cite{BLL004,EY004}. More generally principles of Quantum
Mechanics also are modified to make various problems accessible
to simulations, \cite{BS016}.}

At the same time the idea that modification of the equations
would not affect, at least in some important cases, most of their
predictions arose in other domains: it appeared for instance, in
\cite{SJ993}, to show that the Navier-Stokes (IN above) equation
could be modified, into new reversible equations, still remaining
consistent with selected predictions of the Obukov-Kolmogorov
theory.

In \cite{Ga996b,Ga997b} an attempt was presented to link 
empirical equivalence observations to the well established theory
of the equivalence of ensembles in Statistical Mechanics.
\footnote{\small A naive version would be to claim that
  modifications of equations describing given phenomena will not
  alter 'many other' properties of their solutions if the
  modifications have the effect that properties known to hold, by
  emprical or theoretical analysis, are \ap verified: of course
  the question is 'which are the other properties?' and 'are they
  interesting?'.}  And a paradigmatic example was the NS
incompressible fluid in the simple case of periodic boundary
conditions and forcing acting at large scale (\ie with a force
with Fourier's coefficients non zero only for modes $|\kk|<K_f$
for some $K_f$). In this case new equation proposed was:
\be
  \Dot\uu=-(\T\uu\cdot\T\BDpr)\uu +\a(\uu)\D \uu +\ff -\BDpr
  P\Eq{e3.1}\ee
with the multiplier $\a(\uu)$ so defined that a ``global''
quantity becomes a constant of motion: for instance
the {\it energy} $\EE(\uu)=\sum_{\kk,i}
\frac12|u_{\kk,i}|^2$ or the {\it enstrophy }
$\DD(\uu)=\sum_{\kk,i} \kk^2|u_{\kk,i}|^2$.

In Statistical Mechanics global conserved quantities define the
{\it ensembles}, which are collections of stationary probability
distributions on phase space giving the statistical fluctuations
of observables in the 'equilibrium states'.

The main property being that the ``local'' observables have in
each state properties {\it independent} on the special global
quantity that defines a given state, at least in some limiting
situation (like in the ``thermodynamic limit'', in which the
container volume $\to\infty$).

Distinction between local and global observables is
essential: in particle systems global quantities can be the total
energy (microcanonical ensemble) or the total kinetic energy
(isokinetic ensemble) or the total potential energy or the
average value of certain observables (like the kinetic energy, in
the canonical ensemble).

Local observables, in such systems, are observables $O_{V_0}(\V
q,\V p)$ whose value depends on the configuration of positions
and velocities of particles located, at the time of observation,
in a region $V_0$ of finite size compared to the total volume $V$
of the system.\footnote{\small Always to be thought as $\gg
  V_0$.}

And local observables, in most\footnote{With notable exceptions
like the integrable systems, like the Toda lattice, Calogero lattice...}
systems and in stationary states, evolve exhibiting statistical
properties of the values of $O_{V_0}$ which have a limit as
$V\to\infty$, {\it for all $V_0$}, \ie become independent of the
``volume cut-off $V$''.\footnote{\small {\it Necessary in almost
  all cases} because of lack of existence-uniqueness of solutions
of the equations of motion in infinite volume, just as in the IN
equations in 3D with infinite cut-off.}

Local and global observables arise often also in connection with
the theory of many systems whose evolution is controlled by
differential equations not arising directly from fundamental
equations (like the Lorenz96 model, \cite{GL014}).

In the next section the example of the fluid equations, always in
presence of a UV cut-off $N$, in Eq.\equ{e1.4},\equ{e2.2}) will
be analyzed choosing viscosity force as $\n\D\uu$ or
$\a(\uu)\D\uu$ as in Eq.\equ{e3.1} with:
\be \eqalign{(1)&\quad
  \a(\uu)=\frac{\sum_\kk \ff_\kk\cdot\lis\uu_\kk}
    {\sum_\kk \kk^2|\uu_\kk|^2}\cr
  (2)&\quad\a(\uu)=\frac{\L(\uu)+\sum_\kk \kk^2
    \ff_\kk\cdot\lis\uu_\kk}{\sum_\kk \kk^4|\uu_\kk|^2}\cr}
\Eq{e3.2}\ee
where, with $D$ introduced in Eq.\equ{e2.2}:
\be \L(\uu)=\sum_{\kk_1+\kk_2+\kk_3=0}
D^{i,j,r}_{
  \kk_1,\kk_2,\kk_3}\kk_3^2\, u_{\kk_1,i}u_{\kk_2,j}u_{\kk_3,r}
\Eq{e3.3}
\ee
With the choice (1) the equation Eq.\equ{e3.1} generates
evolutions {\it conserving exactly} the energy $\EE(\uu)$,
considered in \cite{SDNKT018}, while with the choice (2)
evolution {\it conserves exactly} the enstrophy $\DD(\uu)$,
considered in \cite{Ga020,Ga020b}.  But remark that $\L\equiv0$
in 2D, implied by Eq.\equ{e2.3}.

\def\SEC{Ensembles }
\section{\SEC}
\label{sec4}
\iniz

In the case of the fluid equations in Eq.\equ{e1.4} and
Eq.\equ{e3.1} define:
\*
\0{\it Local observables: are functions of the velovity fields
  $\uu$ which depend on the Fourier's modes $\uu_\kk$ with
  $|\kk|<\lis K$ with $\lis K\ll N$, \ie of finite size compared
  to the maximum value $N$ (UV cut-off) used to make the
  equations meaningful}{\ }\footnote{The latter value $N$,
  ``ultraviolet cut-off'', is certainly necessary in 3D $\bf E$,
  \cite{BV019}, just to make sense of the equations, and
  ``might'' be necessary in 3D IN, \cite{Fe000}.}
\*

It can be said that local observables refer to measurements that
can be effected looking at large scale properties of the fluid.

While in Statistical Mechanics locality refers to events in
regions in position space small with respect to the volume
cut-off $V$, in fluid mechanics locality refers to events
measurable in regions in Fourier's space small compared to the
ultraviolet cut-off $N$. Hence locality has a physical meaning
when the aim of the theory is to study properties of ``large
scale'' observables (\ie expressible in terms of Fourier's
components $\uu_\kk$ of the velocity fields $\uu$ with
$|\kk|^{-1}$ of the order of the linear size of the container).

Hereafter consider Eq.\equ{e3.1} and Eq.\equ{e1.4} with $\ff$
fixed and with only few Fourier's components non zero, say
$|\kk|<K_f$ with $K_f$ fixed, and $\|\ff\|_2=1$: such $\ff$ will
be called a ``large scale forcing''.

Having named \equ{e2.2} ``irreversible'' IN, consistently
the Eq.\equ{e3.2} will be named ``reversible'' RE in case (1), or
``reversible'' RN in case (2).  Properties of RE,RN are: \*

\0(1) they generate reversible evolutions $\uu\to S_t \uu$: \ie
if $I\uu=-\uu$ is the ``time reversal'' then $I S_t=S_{-t}I$.  \\
\0(2) RE evolutions conserve exactly energy $E=\EE(\uu)$
and RN conserve exactly enstrophy $D=\DD(\uu)$.
\*

Stationary distributions are usually associated with chaotic
evolutions: therefore the multipliers $\a(\uu(t))$ in
Eq.\equ{e3.1} should show, at large $E$ or $D$, chaotic
fluctuations and behave effectively as a constants: this leads to
several ``equivalence conjectures''.

For clarity we reintroduce a label $N$ as a reminder that all
quantities considered so far were defined in presence of a UV
cut-off $N$ and to discuss variations of $N$.

Collect the invariant (\ie stationary) distributions for IN,RE,RN
and denote the collections $\EE^{IN}_N,\EE^{RE}_N,\EE^{RN}_N$
respectively: we call each such collection an {\it ensemble}.

The stationary distributions are parameterized by the viscosity
$\n$ in $\EE^{IN}$ or by the energy $E$ in the $\EE^{RE}$ or by
the enstrophy $D$ in $\EE^{IN}$.

Denoting as $\m^{IN,\n}_N,\m^{RE,E}_N,\m^{RN,D}_N$ the stationary
distributions, respectively, in the ensembles
$\EE^{IN}_N\EE^{RE}_N,\EE^{RN}_N$, we shall try to establish a
correspondence between the elements $\m^{IN,\n}_N$, $\m^{RE,E}_N$,
$\m^{RN,D}_N$ so that corresponding distributions can be called
``equivalent'' in the sense discussed below.

To fix the ideas we focus first on the correspondence between the
distributions in $\EE^{IN}_N$ and $\EE^{RN}_N$: the simplest
situation arises when above equations, for each $\n$ small or $D$
large, admit a unique stable invariant distribution, \ie a unique
``natural stationary distribution'' in the sense of
\cite{Ru989,Ru995,Ru000b},\footnote{\small {\it I.e.}  almost all
  initial data selected via a probability with continuous density
  $\r(\uu)d\uu$ on the $\NN$-dimensional phase space $\CC_N$, as
  in Sec.\ref{sec2}, assign the same statistics to the
  time-fluctuations of the observables.\label{natural}} a key
concept whose relevance has been stressed since \cite{Ru978b}.

For an observable $O(\uu)$ define
$\media{O}^{IN,\n}_N\defi\m^{IN,\n}_N(O)$,
$\media{O}^{RN,D}_N\defi\m^{RN,D}_N(O)$ the respective time
averages of $O(\uu(t))$ observed under the ($N$-regularized) $IN$
and $RN$ evolutions.

Define also the {\it work per unit time} done by the forcing:
\be L(\uu)=\int_{\TT^d} \ff(x)\cdot\uu(x)\frac{dx}{(2
  \p)^d}=
\sum_\kk \ff_\kk\cdot\lis\uu_\kk,\Eq{e4.1}\ee
So the average work per unit time in the stationary states with
parameters $\n$ or $D$ of the ensembles $\EE^{IN}_N,\EE^{RN}_N$
is, respectively, $\media{L}^{IN,\n}_N\equiv \m^{IN,\n}_N(L)$ or
$\media{L}^{RN,D}_N\equiv \m^{RN,D}_N(L)$.

Given $\m^{RN,D}_N, \m^{IN,\n}_N$: define $\m^{IN,\n}_N$ to be
{\it correspondent} to $\m^{RN,D}_N$, denote this by
$\m^{IN,\n}_N\sim\m^{RN,D}_N $, if the time average of the {\it
  enstrophy } is equal in the two distributions:
\be \media{\DD}^{IN,\n}_N=D\Eq{e4.2}\ee
The natural distributions, see footnote\ref{natural}, are
associated with chaotic evolutions: therefore the multipliers
$\a(\uu(t))$ should show, at large $D$, chaotic fluctuations and
behave effectively as constants equal to their average.

Hence the proposal, \cite{Ga996b,Ga997b}: for an observable
$O(\uu)$ define $\media{O}^N_{\n}\defi\m^{IN,\n}_N(O)$,
$\media{O}^N_{D}\defi\m^{RN,D}(O)$ the respective time averages
of $O(\uu(t))$ observed under the IN and RN
evolutions; then:
\*

\0{\rm Conjecture 1:} {\it Under the equivalence condition
  Eq.\equ{e4.2}, equal average enstrophy, if $O(\uu)$ is an
  observable, then:}
\be\lim_{\n\to0} \media{O}^N_{\n}=
\lim_{\n\to0}\media{O}^N_D\Eq{e4.3}\ee
\*

The collection of stationary distributions $\m\in\EE^{IN}$ can be
assimilated to the distributions of Statistical Mechanics
canonical ensemble and the distributions $\m\in\EE^{RN}$ can be
assimilated to the distributions of the microcanonical
ensemble. The regularization $N$ plays the role of the volume and
the friction $\n$ that of temperature, the enstrophy that of
energy.

So there is ´some´ similarity between the equilibrium states
equivalence in Statistical Mechanics and the equivalence proposed
by the conjecture 1 about averages observed following the two
different evolutions $IN$ and $RN$, under the condition of equal
average enstrophy.

\def\SEC{Ensembles in Fluids and Statistical Mechanics}
\section{\SEC}
\label{sec5}
\iniz

However the need to consider the limit as $\n\to0$ in
Eq.\equ{e4.2} limits strongly the analogy: the Statistical
Mechanics theory of equivalence of the ensembles requires
considering the thermodynamic limit $V\to\infty$ of the volume of
the system container and {\it restricting} the observables $O$ to
be {\it local}.

In the conjecture in Sec.\ref{sec4}, instead, the observables are
{\it unrestricted} and the role of the volume $V$ is played by
the cut-off $N$. Clearly for a full analogy equivalence should
hold for {\it $\n$ fixed} as $N\to\infty$, provided the
observables are suitably restricted.\footnote{\small For instance
  in Statistical Mechanics microcanonical and canonical ensembles
  are equivalent unless, of course, one is interested in the
  fluctuations of the (global) observable 'total energy'.}

To see what has to be understood to try to establish
a closer connection between the theory of the ensembles in
Statistical Mechanics and the proposed fluid equations
equivalence the key remark is that the conjectured equivalence is
based on the chaoticity of the evolution, which is ensured by the
$\n\to0$ condition in Eq.\equ{e4.2}. 

So the same argument can simply be extended to many other
equations in which the size of a parameter controls the
increasingly ``chaotic'' motion of a system. Examples of this
phenomenon have been explicitly considered adding new examples to
a wide literature of {\it homogenization} phenomena: see
\cite{GRS004,Ga018,Ga019c,JC020} for fluid equations or
\cite{GL014,Ga019c,BCDGL018}. Thus the conjecture in
Sec.\ref{sec4} although quite unsatisfactory, as pointed out,
seems to hold in its generality, \cite{GRS004,GL014}.

{\it Far more interesting} would be to dispose of the condition
$\n\to0$ and to realize a stronger analogy with Statistical
Mechanics. The idea is that some, by far not all, equations
describing macroscopic phenomena arise as scalingn limits of
fundamental equations governing evolutions of systems of
particles interacting via forces veryfying all principles and
symmetries of Physics: staying within classical systems among
these are Newton's laws, time reversal and parity and charge
symmetry...

The evolutions, at so fundamental a level, are certainly chaotic
and the ergodic hypothesis epitomizes this property: from them,
via approximations and/or heuristic arguments, arise simplified
equations ({\it models}) that generate motions apt to describe
many of the features found in the observations. One of the first
examples is in the derivation of the (compressible) Navier-Stokes
equations in \cite{Ma867-b}.

A model can even fail to respect one or more of the fundamental
laws or symmetries: like the time reversal symmetry breaking
which accounts phenomenologically for dissipation. This has never
been considered a violation of the basic principles: it has been
always clear that it was simply due to the procedure followed in
the derivations.

Then the idea arises that there could (should ?) exist models
representing the same phenomena at the same level of accuracy and
preserving some of the properties that other models do not
respect, but which are properties on which there is a minor
interest in the context on which one is working,
\cite{Ga020,Ga020b}.

The case of the Navier-Stokes equation has been proposed as an
example of the possibility of describing an incompressible fluid
via a reversible equation, without the need (as in conjecture 1
above, see also \cite{Ga013b}) of taking the limit $\n\to0$ but
paying the price of restricting attention to a suitable (large)
family of observables.

In the NS case the equations of motion are irreversible but they
arise from a fundamental microscopic representation which is
reversible and chaotic.  If, as in most experimental studies,
interest is on properties of ``large scale'' then it is natural
to extend the conjecture 1 to the NS evolution {\it without
  cut-off} restricting
attention to the case in which the macroscopic forces act at
large scale and whose results have to be observed also on large
scale, \cite{Ga020,Ga020b}.

This can be formalized, \cite{Ga020b}, into the: \*

\0{\rm Conjecture 2:} {\it Under the equal average enstrophy
  Eq.\equ{e4.2} and if $O$ is a local observable, as defined in
  Sec.\ref{sec4}, then
\be\lim_{N\to\infty} \media{O}^N_{\n}=
\lim_{N\to\infty}\media{O}^N_D\Eq{e5.1}\ee
for all $\n>0$.}
\*

The conjecture 2 therefore adds to conjecture 1 the restriction
that the observables $O$ {\it must be local} and replaces the
equivalence condition $\n\to0$ with the {\it condition
  $N\to\infty$} (keeping equal average enstrophy).  

The ensemble $\EE^{IN,\n}_N$ is analogous to the canonical
ensemble with $\n$ as temperature while $\EE^{IN,D}_N$ is
analogous to the microcanonical ensemble with the enstrophy $D$
as the energy and $N\to\infty$ corresponds to $V\to\infty$, \ie
to the thermodynamic limit necessary for all local observables to
show the same statistics.

The analogy with Statistical Mechanics is now 'essentially'
complete (however see Sec.\ref{sec6} below) and provides an
example of use of the 'thermodynamic limit' among the ideas
emerging in nonequilibrium theory, \cite{Ru999,AB020}.

\def\SEC{Equivalence and phase transitions}
\section{\SEC}
\label{sec6}
\iniz

Conjecture 2 of Sec.\ref{sec5} leaves a gap in the strict analogy
between Fluid Mechanics and Statistical Mechanics ensembles. Is
there an analogue of the phase transitions?

So far we have considered the ensembles $\EE^{IN,\n},\EE^{RN,D}$
assuming that for each pair of $\n,D$ the equations $IN$ and $RN$
admit just one ``natural'' stationary distribution controlling
the fluctuations of the (local) observables.

However it is possible that initial data chosen with a
distribution density $\r(\uu)>0$ generate a statistics which
still depends on the initial $\uu$ with positive probability:
this case would be met if the evolution admitted several
attracting sets in the phase space $\CC_\NN$. 

If so, label the
``indecomposable'' invariant distributions by $\m_\th\in
\EE^{IN,\n}_N$, $ \th=1,2,\ldots,q_{\n,N}$.\footnote{\small
  Indecomposable means that for each $\th$ with probabiity $1$
  with respect to $\m_\th$ initial data generate precisely
  $\m_\th$ itself: synonimous of ergodic.} Likewise label the
``indecomposable'' invariant distributions by $\m_\th\in
\EE^{RN,D}_N$, $ \th=1,2,\ldots, p_{D,N}$. Each $\m_\th$ will be
called a ``pure phase``.

For simplicity we assume that $q_\n,p_D<\infty$ and say that at
the values $\n$ or $D$ there are $q_\n$ or $p_D$ ``pure phases''.

Then, keeping in mind the theory of phase transitins in
Statistical Mechanics, conjecture 2 should be modified as:
\*

\0{\it If under the equivalence condition between $\n$ and $D$,
  Eq.\equ{e4.2}, there are $q_{\n,N}$ respectively $p_{D,N}$ pure
  phases, then $q_{\n,N},p_{D,N}$ have the same limit $q\ge1$ as
  $N\to\infty$, and it is possible to establish a $1\otto1$
  correspondence between the $\m_j\in \EE^{RN,\n}_N$ and the
  $\m_j\in \EE^{RN,D}_N$ such that the distribution of the local
  observables become, in corresponding $\m$'s and in the
  limit $N\to\infty$, the same.}
\*

If one thinks to the ferromagnetic Ising model in voume $V$ at
low temperature then there are two indecomposable pure phases in
which the total magnetization or just its average is fixed to
some $m=\pm m^*\ne0$, \cite{GM972a}, whether the boundary
conditions are periodic or free or whether the dynamics is of
Glauber type or other. Make correspondent the phases with the
same $m$ then the local observables (\ie the observables $O$
which depend only on the spins located in a fixed region) have
fluctuation with the same statistics in the thermodynamic limit,
$V\to\infty$.

See comments following Eq.\equ{e9.1} for other analogies with
phase transitions arising in $RN$ and deveoped in
\cite{SDNKT018}.

\def\SEC{Chaotic hypothesis and reversibility}
\section{\SEC}
\label{sec7}
\iniz

In a general evolution equation $\dot x= g(x), x\in bR^n$
generating motions $t\to S_tx$ which lead to an attracting
set\footnote{\small This is a compact set $\AA$ such that all
  initial data $x$ close enough to $\AA$ are such that the
  distance $d(S_tx,\AA)\tende{t\to\infty}0$.}  $\AA$ on which
they are chaotic (\ie have positive Lyapunov exponents) the
``chaotic hypothesis'' is: \*

\0{Chaotic hypothesis (CH): \it The attracting sets can be
  considered smooth surfaces on which the motion is an Anosov
  flow, \cite{GC995b,Ga995b}.\footnote{\small Anosov evolutions
    are smooth flows on bounded smooth surfaces $\AA$ such that
    at every point $x$ the evolution is hyperbolic (\ie in a
    system of coordinates following $S_tx$ as $t$ varies the
    $S_tx$ is a hyperbolic fixed point); furthermore any open set
    $U\subset\AA$ is such $S_tU$ covers any prefixed point
    $x\in\AA$ for infinitely many $t>t_0 $ and for all $t_0$
    (``motion of most points covers densely $\AA$'',
    ``recurrence''), \cite{AA966,HK995}.}} \*

The assumption implies the exsistence of a unique stationary
probability distribution $\m$ on $\AA$ which is a natural
distribution in the sense that it gives the statistical
properties of the motions of almost all initial data chosen in
the vicinity of $\AA$ with a probability with density $\r(x)>0$.

This assumption should be viewed as an extension of the analysis
leading to the ergodic hypothesis in equilibrium problems,
although of course examples which do not satisfy it are easy to
find.

Still it is an assumption that has been proposed to be applicable
to most systems undergoing chaotic motions, following a path that
led to the modern ergodic hypothesis in equilibrium
thermodynamics. In the case of non equilibrium the chaotic
hypothesis only strengthens the key ideas of Ruelle,
\cite{Ru989,Ru978b,Ru995,Ru999}, developed to provide a
fundamental reason for the selection of the probability
distributions to be used to evaluate the time averages of
observables in systems out of equilibrium, remaining compatible
with leading to the selection of the microcanonical ensemble in
the equilibrium cases.

The real problem is to show that it not only has the merit of
providing a conceptual extension of ideas at the basis of
equilibrium Statistical Mechanics to nonequilibrium and Fluid
Mechanics but it has also predictive power on new observations. 

The simplest applications of the CH deal with reversible
evolutions; hence the equations $RN$ or $RE$ might offer 
insights.

Imagine to fix the UV cut-off $N$ and that for some $\n$ the
evolution appears to generate trajectories of $IN$ that visit
densely the entire phase space. We expect that to be the case at
small $\n$, at fixed $N$: and for $\n=0$ ergodicity is expected
to hold. As $\n$ increases the system develops an attracting set
which, if the CH holds, should still be the full phase space (a
consequence of the structural stability of Anosov
systems\footnote{\small Structural stability means here that
  small pertubations of Anosov systems are still Anosov
  systems.\cite{AS967,Si968a,Ru010}.}).

For such value of $\n$ let $D$ be the average enstrophy: we
consider the $RN$ evoution of initial data with enstrophy
$\DD(\uu)=D$. The phase space ``contracts'' at a rate $\s(\uu)$,
\ie. if $u_{\b,\kk}=u_{r,\b,\kk}+i u_{i,\b,\kk},\,\b=1,2$, see
\Eq.\equ{e2.1}, at a rate equal (by Liouville's theorem) to:
\be -\s(\uu)=-\sum_{\kk,\b}^*
\Big(\frac{\dpr\dot u_{r,\kk,\b}}{\dpr u_{r,\kk,\b}}+
\frac{\dpr\dot u_{i,\kk,\b}}{\dpr u_{i,\kk,\b}}\Big)
\Eq{e7.1}\ee
where $\sum^*_\kk$ denotes
summation over the $\kk$ so that only one $\kk$ between $\pm\kk$
contributes (the contribution is independent on which one is
selected).

Let
$F_4=\sum^*_\kk \kk^4 \ff_\kk\lis u_\kk$,
$E_6=\sum_\kk^*\kk^6|\uu_\kk|^2$, $E_4=\sum^*_\kk \kk^4
|\uu_\kk|^2$, $K_2=\sum^*_\kk \kk^2$,  then:
\be
-\s(\uu)=2\Big(K_2-\frac{E_6(\uu)}{E_4(\uu)}\Big)\,\a(\uu)+
\frac{F_4(\uu)}{E_4(\uu)}
\Eq{e7.2}\ee
which has the same expression in dimension $2,3$ (but the
expression of $\a$ is of course different). 

If CH holds the ``Fluctuation theorem'', FT, can be applied and
the result is that it implies a simple prediction on the {\it non
  local} observable
\be p=\frac1t \int_0^t \frac{\s(\uu(t'))}{\s_+}dt'\Eq{e7.3}\ee
where $\s_+$ is the average value of $\s(\uu(t))$.
The fluctuations of $p$ in the stationary distribution
$\m^{RN,D}_N$ have the probability that $p\in [a,b]$
is $\exp\,(t\max_{p\in [a,b]} s(p) + o(t))$ and the ``large
deviations rate'' $s(p)$ has the symmetry property,
\cite{GC995,Ga995b,GC995b}:
\be s(p)-s(-p)= p\,t\,\s_+\Eq{e7.4}\ee
which follows combining CH and the time reversibility.

{\it The observable $\s(\uu)$ can be considered also as an
  observable for the {\rm IN} evolution}. Although it is non
local it has been tested in a few cases to see whether it
nevertheless obeys the same fluctuation relation Eq.\equ{e7.4} in
corresponding distributions, see \cite{Ga020} for a positive
result, but no other attempt exist, so far, to check possible
equivalence between corresponding fluctuation relations (in any
event the fluctuation relation is not a local observable and is
not covered by the conjectures.

\def\SEC{Attractors and small scales}
\section{\SEC}
\label{sec8}
\iniz

However the assumption that at an enstrophy value $D$ the
stationary distribution $\m^{RN,D}_N$ arises from
an evolution which leads to an attracting set invariant under
time reversal is too strong.

Certainly it does not cover the cases in which the UV cut-off $N$
is large enough and the $\uu_\kk$ components are strongly damped
for $|\kk|$ large (as implied by the equivalence conjecture).

Hence if $N$ is large the attracting set $\AA$ will shrink and
its time reversal image $I\AA$ will become different from $\AA$:
a {\it spontaneous breaking of time reversal}.

The consequence is that the FT cannot be applied to the
observable $\s(\uu)$, not even if the CH is assumed in the
reversible RN equation.

Nevertheless FT could be applied, under the CH, to the motion on
$\AA$ {\bf if} the time reversal $I$ could be replaced by {\it
  another map} $\wt I$ which leaves $\AA$ invariant and on $\AA$
the $\wt I S_t=S_{-t} \wt I$ holds. Because by CH $\AA$ is a
surface on which the evolution is of Anosov type.

In this case the fluctuation relation will be applied no longer
to $\s(\uu)$, but to the sum $\s_\AA$ of the {\it local Lyapunov
  exponents}
\footnote{\small The local exponents are defined as the
  eigenvalues of the symmetric part of the Jacobian of the motion
  on $\AA$: their sum defines the contraction (or expansion) of
  the surface elements of $\AA$.} relative to the motion on
$\AA$: clearly the negative exponents pertaining to the
attraction to $\AA$ {\it should not} be counted.

Hence the question under which conditions a time
reversal for the motions on $\AA$ exists is preliminary to the
second question of how to identify the Lyapunov exponents of the
motions on $\AA$.

Considering the RN equations with UV cut-off $N$ and fixed
enstrophy $D$. Suppose that for small $N$ (\ie at strong
regularization) the motions invade densely the phase space
$\CC_N$: \ie the attracting set $\AA$ coincides with
$\CC_N$. Increasing $N$ arrives a $N_c$ beyond which the
(average) viscosity affects the components $u_\kk$ with large
$\kk$ so that $\AA$ becomes smaller than $\CC_N$.

So the evolution is reversible for all $N$, but for $N$ large its
restriction to the attracting set $\AA$ is not.

In \cite{BG997} the question of existence, as a ``remnant'' of
the global symmetry $I$, of a time reversal $\wt I$ operating on
$\AA$ has been examined and a geometric property, named {\it
  Axiom C} property, leading to the existence of $\wt I$ was
identified and shown to have a ``structural stability'' property
(as demanded to properties of physical
relevance).\footnote{\small \ie small pertubations of systems
  with the axiom C property still have the property,
  \cite{BG997}. Persistence under perturbations is clearly
  essential in most Physics theories, \cite{Ru977}.}
The definition and main properties of Axiom C are described in
Appendix \ref{secd}.

A scenario for the application to IN,RN (and more general)
equations in which time reversal is a symmetry but $\AA$ does not
coincide with the full phase space can be the following.

Assume that Axiom C holds for RN, hence there is a map $\wt
I:\AA\to\AA$ such that $\wt I S_t=S_{-t}\wt I$: to apply FT the
problem still remains of identifying the phase space contraction
$\s_\AA$, \ie the local Lyapunov exponents which contribute to
the phase space contraction on the surface $\AA$.

In studying the Lyapunov spectrum for $IN,RN$ the following
``pairing symmetry'' has been tested and {\it approximately}
verified in a {\bf few} 2D simulations and for a few values of
the ensembles parameters $\n,D$.

If the $\NN$ {\it local} Lyapunov exponents are arranged
indecreasing order and their time averages are $\l_0\ge \l_1\ge
\ldots,\ge \l_{\NN-1}$, then
\be (\l_k+\l_{\NN-1-k})=n+O(k^{-1}),\
k=0,\ldots,\frac\NN2 \Eq{e8.1}\ee
and the constant $n<0$ and the $\l_k$ turned out to have, for
each $k$, in IN and RN very different fluctuations but remarkably
the {\it same average} in corresponding distributions
$\m^{IN,\n}_N$ and $\m^{RN,D}_N$: quite unexpected a result
because the $\l_k$ are not local observables. See figs.7,8 and,
respectively, figs.5,6 in \cite{Ga020,Ga020b}. Relations like
Eq.\equ{e8.1} are called ``pairing rules''.

So among the $\NN/2$ averages $\l_k,\l_{\NN-1-k}$ there may be,
depending on the values of $\n,D$, pairs in which both elements
are $<0$: \ie there may be $n^*\le \NN/2$ pairs of opposite sign
and $\NN/2 -n^*$ negative pairs.

A natural interpretation of the above pairing rule is that the
pairs of exponents $<0$ represent the exponents controlling the
approach to $\AA$ while the other $n^*$ pairs are associated with
the chaotic motion on the attracting set; the phase space
contraction on $\AA$ would then be $\s_\AA(\uu)=\sum_{k=0}^{n^*}(
\l_k(\uu)+\l_{\NN-1-k}(\uu))$.

The interest of the above remarks is that {\bf if} CH, axiom C
and pairing are satisfied and if the $O(k^{-1})$ in Eq.\equ{e8.1}
can be neglected the consequent relation:
\be \s_\AA(\uu)=\frac{2n^*}{\NN}\s(\uu)\Eq{e8.2}\ee
can be used to define the phase space contraction on $\AA$.  The
advantage is that $\s_\AA$ is measurable simply by measuring
$\s(\uu)$ from the equations of motion using
Eq.\equ{e8.2}.\footnote{Remark that this evaluation of the
attracting set dimension is different from that obtained by the
general Kaplan-Yorke dimension: an upper bound on the latter is, for IN,
in \cite{Ru982}.}

Then, applying FT, the relation Eq.\equ{e7.4} is simply changed
into:
\be s(p)-s(-p)= p\,t\,\frac{2n^*}{\NN}\,\s_+\Eq{e8.3}\ee
in the case of the RN evolution.

Furthermore {\it if} the equivalence conjecture can be {\it
  extended} to the non local observables $\s,\s_\AA$ then the
fluctuation relation gives a prediction on fluctuations of both
IN and RN and, if $n^*<\NN/2$, a test of the Axiom C.

The above scenario, proposed first in \cite[p.445]{BGG997} and
leading to the formulation of Axiom C, does not seem to have been
tested, not even for simple test examples and it is certainly
interesting if it can be confirmed in some instances: the only
attempt to check Eq.\equ{e8.3} dealt, \cite{Ga020}, with cases in
which $n^*=\NN/2$. Hence it does not deal with the most
interesting part of the above scenario and in particular it does
not test the Axiom C: however it did yield the result that the
fluctuation relation holds in equivalent distributions, \ie the
observable $\s(\uu)$, Eq.\equ{e7.2}, satisfies the same
Eq.\equ{e7.4} even in the irreversible evolution IN.

There are cases in which the phase space contraction can be
identified with entropy creation: this is important as the
entropy production is accessible, in a laboratory experiment, to
measurements of heat and work exchanges with the surroundings,
\cite{Ci017}: however it is very difficult to perform complete
analysis of such energy exchanges and among the many experimental
works very few convincingly discuss the problem.

\def\SEC{Other ensembles}
\section{\SEC}
\label{sec9}
\iniz

In statistical Mechanics there are several equivalent
ensembles. The same should hold for the fluids considered above.
For instance we could compare IN with the equation that will be
called RE given by Eq.\equ{e3.1} with $a(\uu)$ given by the first
of Eq.\equ{e3.2}.

The RE is reversible and conserves the global quantity
$\EE(\uu)$, energy, instead of enstrophy. The ensemble is now the
collection of the stationary states $\m^{RE,E}_N \in \EE^{RE}_N$.

The equivalence condition is equality of the average
energy, hence Eq.\equ{e4.2} is modified as
\be \media{\EE}^{IN,\n}_N=E \Eq{e9.1}\ee
and the analysis of the previous sections can be repeated. 

Care has to be exercized because the condition Eq.\equ{e4.2} is
not the same as $E=\media{\EE}^{IN,\n}_N$ (unlike the
corresponding case of the RN equations).\footnote{\small If the
  RN equation is multiplied side by side by $\lis u_{\b,\kk}$ and
  the result is summed ove $\kk,\b$, it immediately follows that
  both conjectures imply $D=\media{\DD}^{IN,\n}_N$. And
  conversely the latter equality implies the equal dissipation
  property, Eq.\equ{e4.2}, hence the condition for
  equivalence.}${}^,$\footnote{\small The ensemble $\EE^{RE}_N$,
  in which the global quantity conserved is the energy rather
  than the enstrophy, has been considered in detail in
  \cite{SDNKT018} where a different kind of very interesting
  phase transition phenomena occurring in the RE equations is
  studied. In the limiting case in which $\n\to0$ as well as in
  the analysis of the transition in \cite{SDNKT018} it is likely
  that the difference bewteen imposing the condition
  $E=\media{\EE}^{IN,\n}_N$ instead of the equal average
  enstrophy, as in Eq.\equ{e9.1}, is not appreciable.}

This implies that a first test of the
conjecture in the case of RN is obtained by fixing $\n$ and
computing the average enstrophy $D$ and checking that if $\n,D$
correspond in the sense of Eq.\equ{e4.2}:

\be \media{\a }^{RN,D}_N\,D=\n\media{\DD}^{IN,\n}_N, \quad\ie\
\media{\a}^{RN,N}=\n\Eq{e9.2}\ee
where $\DD(\uu)$ denotes, as above, the enstrophy.  While for RE,
if $\n$ and $E$ correspond in the sense of Eq.\equ{e9.1}, the
analogous test is to check
\be \frac{\media{ \a\DD}^{RE,E}}
         {\media{\DD}^{RN,\n}  }
    =\n \Eq{e9.3}\ee
The above relations have been tested in several cases, with
particular care and a few positive results for the equivalence
between IN and RN in 2D;  only in very few cases for the IN and RE
equivalence.

\APPENDICE{1} \def\SEC{Euler flow is geodesic}
\section{\SEC}
\label{seca}
\iniz

Here some details on the Hamiltonian representation Eq.\equ{e1.2}
for the Euler flow are presented, listing again for the reader´s
convenience, the conventions set in Sec.\ref{sec1}.

It has to be kept in mind that in analytic mechanics the
canonical coordinates for $n$-degrees of freedom systems are
given as strings of $2n$ variables $(\{p_i,q_i\}_{i=1}^n)$:
particle $i$ is located at position $q_i$ and has momentum $p_i$.

In a Lagrangian description of a fluid, coordinates will be
$(\qq,\Dot\qq)= (\{q_\x,\dot q_\x\}_{\x\in\TT^d})$ with
$(\qq,\Dot\qq)$ consisting in a diffeomorphism $\qq\,:\,\x\to
q_\x$ in the space $Dif(\TT^d)$ of $\CC^\infty$ diffeomorphisms
of $\TT^d$ and $\Dot\qq \in Lin(\TT^d)$ where $Lin(\TT^d)$ is the
space of the $\CC^\infty$ vector fields 'tangent' to $\qq$: in a
pair $(\qq,\Dot\qq)$ the vector $\dot q_\x\in R^d$ is considered
a vector applied at the point $q_\x$.

Hence, given $\qq,\Dot\qq$, the derivative $\dpr_{q_{\x,i}}\dot
q_{\x,i}$ is defined as well as the divergence $({\rm
  div}\Dot\qq)_\x\defi \sum_{i=1}^d\dpr_{q_{\x,i}}\dot q_{\x,i}$
of $\dot q_\x$.

The space of the pairs $(\qq,\Dot\qq)$ will be called $\FF$ and
the points of $\TT^d$ become labels of a fluid element located at
the point $q_\x$ with velocity $\dot q_\x$.

More formally $(\qq,\Dot\qq)\in Dif(\TT^d)\times Lin(\TT^d)\defi
\FF$ where $Dif(\TT^d)$ is the space of the $\CC^\infty$
diffeomorphisms of $\TT^d$ and $Lin(\TT^d)$ the space of the
$\CC^\infty$ vector fields with $0$ average: for each
$(\qq,\Dot\qq)$ the vector $\dot q_\x\in R^d$ is considered
applied to the point $q_\x$.  and $({\rm
  div}\Dot\qq)_\x=\sum_{i=1}^d\dpr_{q_\x}\dot q_\x$.

Actually we concentrate on the subspace of $(\qq,\Dot\qq)\in
SDif(\TT^d)\times (SLin(\TT^d))\defi S\FF\subset\FF$ where the
evolution of an {\it incompressible} fluid takes place:
$SDif(\TT^d)$ being the {\it volume preserving} diffeomphisms and
$SLin(\TT^d)$ the {\it $0$-divergence} vector fields, \ie for
each such pair $(\qq,\Dot\qq)$ it is $({\rm div}\Dot\qq)_\x=0$.

If the positions $q_\x$ are moved the variation of $\dot q_\x$ is
proportional to $\frac{\dpr \dot q_{\x}}{\dpr{q_{\h}}}$; the
Lagrangian is:
\be \LL(\Dot \qq,\qq)=
\int_{\TT^d} \frac12\Big( \dot q_\x^2
-Q(\Dot \qq,\qq)_\x\Big)d\x
\Eq{a.1}\ee
where $Q$ is the quadratic form on $\FF$:
\be- \frac1{4\p}\int_{\TT^d} d\g\sum_{r,s=1}^d
\Big(
  \frac{1}{''|q_\x-q_\g|''} \Big)\,
  \frac{\dpr \dot q_{\g,r}}{\dpr q_{\g,s}}\,
  \frac{\dpr \dot q_{\g,s}}{\dpr q_{\g,r}}
\Eq{a.2}\ee

\kern-3mm \0and ${-(4\p)^{-1}}/{''|x-y|''}$ symbolizes the
Green's function for the Laplacian on $\TT^3$: \footnote{\ie
  formally the summation over images $y+2\p \V n,\,
  \V n\in Z^d$; which makes sense if the kernel is applied to a
  smooth function with $0$ average.}  so that $\D_{q_\x} Q_\x=
(\T\BDpr \pp)_\x (\BDpr \T\pp)_\x$. Hence $\LL$ is a metric over
$\FF$ and the flow
\vskip-2mm
\0it generates runs over its geodesics.

Addition of $Q$ corresponds to a force which has the property
that it keeps data in $S\FF$ inside $S\FF$ as long as they evolve
smoothly in time: this is checked in the following.

The Lagrangian $\LL$ leads to the canonical variables via:
\be \eqalign{
p_{\x,i}\defi&
\dot q_{\x,i}-\frac{1}{4\p}
\int_{\TT^d}d\l \int_{\TT^d}d\g \frac{1}{\{''|q_\l-q_\g|''\}}\cr
&\cdot\d_{i,r} \frac{\dpr\d(q_\x-q_\g)}{\dpr q_{\g,s}}
\frac{\dpr \dot q_{\g,s}}{\dpr q_{\g,r}}\defi
\dot q_{\x,i}+A(\qq,\Dot q)_{\x,i}\cr}
\Eq{a.3}\ee
where in the last equality (defining $A$) $p_{\x,i}=\dot
q_{\x,i}$ and $A=0$ hold if $\qq\in SDif(\TT^d)$ and $\Dot\qq\in
SLin(\TT^d)$, \ie if $\sum_{s=1}^d\frac{\dpr \dot q_{\g,s}}{\dpr
  q_{\g,s}}=0$: so that the diffeomorphism $\qq$ is
incompressible, and also $\dot\qq$ is divergenceless, ${\rm
  div}\Dot\qq=0$, and the double integral vanishes because $\int
d\l \{''|q_\l-q_\g|''\}^{-1}$ is a constant and integration by
parts over $\g$ becomes possible as $d\g=d q_\g$ if $\qq\in
SDif(\TT^d)$.

Defining the linear operator $A^{-1}(\qq,\pp)$, obtained by
inverting $\pp=\Dot\qq+A(\qq,\Dot q)$, the equation for $\Dot\pp$
is readily obtained, at least if $(\pp,\qq)\in S\FF$:

\be\dot q_\x=p_\x,\qquad \dot p_\x=\dpr_{q_\x}\LL(\Dot q,\qq)=
\dpr_{q_\x} \frac12Q(\Dot\qq,\qq)\Eq{a.4}\ee
and the Hamiltonian:
\be
H(\pp,\qq)\defi \int_{\TT^d}(
\pp_\x\cdot\dot\qq_\x-\LL(\pp-A^{-1}(\qq,\pp),\qq)_\x)d\x
\Eq{a.5}\ee
yields canonical
equations, which {\it for data in $S\FF$ are}:
\be\eqalign{
  &\dot q_\x=p_\x, \quad \dot p_\x= -\dpr_{q_\x} P_\x(\qq,\pp)\cr
& P_\x\defi 
\int_{\TT^d} d\g\sum_{r,s=1}^d
\Big(
  \frac{-(4\p)^{-1}}{''|q_\x-q_\g|''} \Big)\,
  \frac{\dpr p_{\g,r}}{\dpr q_{\g,s}}\,
  \frac{\dpr p_{\g,s}}{\dpr q_{\g,r}}\cr}\Eq{a.6}\ee
which hold only if $(\pp,\qq)\in S\FF$, while in $\FF$ the
equations would be more involved (but uninteresting for the
present purposes) although still Hamiltonian.

The equations can be written {\it for data in} $S\FF$ setting
$x=q_\x, u(x)=p_\x, P(x)= P_\x$. Then $\frac{d
  p_{\x,i}}{dt}=\dpr_t p_{\x,i}+ \sum_j
p_{\x,j}\dpr_{q_{\x,j}}p_{\x,i}$ and:
\be\kern-3mm \eqalign{
  &{\BDpr\cdot u}(x)=0,\ P(x)=
  -\kern-1mm\sum_{r,s=1}^d\kern-1mm
  \D^{-1}(\dpr_s u_r(x) \dpr_r u_s(x))\cr &
  \dpr_t q_\x=u(q_\x),\quad \dpr_tu( x)=
  -\T u( x)\cdot\T\BDpr u( x)-\BDpr P( x) \cr
  &\kern3.8cm+\ff( x)+\n\D u( x)\cr}\Eq{a.7}\ee
where the terms in the third line are added in the case there is
forcing and NS viscosity.

Therefore the Eq.\equ{a.7} coincide with the Navier Stokes
equations and their solutions will remain in $S\FF$ s long as
they remain smooth: for data not in $S\FF$ only solutions local
in time can be envisaged and the equations would be more
involved.\footnote{\small Representing a constrained motion as a
  special case of unconstrained motion subject suitable extra
  forces follows a familiar prototype. A point mass constrained
  on a circle of radius $R$, centered at the origin $O\in R^3$,
  can be seen as a point subject to a centripetal force evolving
  under the Lagrangian $\LL=\frac12
  {\Dot\qq^2}-\frac{(|\qq|-R)}R{\Dot\qq^2}$.
This leads to $\pp=\Dot\qq \th$ with $\th=(1-2\frac{|\qq|-R}R)$
and, for the Hamiltonian, $\HH=\frac12 \frac{\pp^2}{\th}$, to the
equations $\Dot\qq=\pp\th^{-1},\ \Dot\pp=-\frac{\pp^2}{\th^2
  R}\frac{\qq}{|\qq|}$.  Thus it appears that the phase
space $R^6$ is analogous to $\FF$, the maps of the circle $\x\to
s$, mapping the arc $\x$ to the arc $s$, correspond with $SDif$
and the vectors tangent to the circle are analogous to
$SLin$. The motion is in general a geodesic motion, as long as it
is defined (\ie as long as $|\qq|\ne0,\infty$), which for data
initially on the circle and initial velocity tangent to it is a
uniform rotation.  On such motions the Hamiltonian value is
$\frac12\pp^2$ as $\th=1$, alike the corresponding vanishing
of $Q$ on $S\FF$ in Eq.\equ{e1.3},\equ{a.5}.}

\APPENDICE{2}
\def\SEC{Euler´s equation Jacobian}
\section{\SEC}
\label{secb}
\iniz

The Jacobian is obtained by taking suitable functional
derivatives of the transport term and the pressure term, before
applying the projection operator $\cal P$ in Eq.\equ{e2.4}. The
contribution of the transport term is (before applying $\cal P$):
\be\frac{\dpr \dot u_i(x) }{\dpr u_j(y) }
=-\d(x-y) \dpr_{x_j} u_i(x)-u_k(x)\d_{i,j}\dpr_{x_k}\d(x-y) 
\Eq{b.1}\ee
where the second term is an {\it antisymmetric} operator in
$L_2(\TT^d)\times R^d$.
The contribution from the pressure term
is
\be \eqalign{\frac {\dpr\ \dpr_{x_i} P(x)} {\dpr
    u_{j}(y)}=&-2\dpr_{x_i} \int_{\TT^d}dz\,\D^{-1}(x-z)\cr&\cdot
  \dpr_{z_k}\d(z-y) \dpr_{z_j}u_k(z)\cr
  =&2(\dpr_{x_i}\dpr_{y_k}\D^{-1}(x-y))\dpr_{y_j} u_k(y)\cr
}\Eq{b.2} \ee
and in both Eq.\equ{b.1},\equ{b.2} summation over $k$ is
intended.

The latter operator does not contribute to the Jacobian
because acting on a divergenceless field yields $0$; therefore
the {\it symmetric part} of the Jacobian is the multiplication
operator, in $L_2(\TT^d)\times R^d$, by:
\be W_{i,j}(x)=-\frac12(\dpr_{x_j} u_i(x)+\dpr_{x_j}
u_i(x))\Eq{b.3}\ee
followed by the orthogonal projection $\cal P$ on the subspace of
the divergenceless fields $SLin(\TT^d)\subset L_2(\TT^d)\times
R^d$, Sec.\ref{sec1}.

\APPENDICE{3}
\def\SEC{The Axiom C.}
\section{\SEC}
\label{secd}
\iniz

To describe the main features of the Axiom C, \cite{BG997},
consider first the simpler case of a {\it reversible
  diffeomorphism} $S$, \ie such that there is a diffeomorphism,
$I$ such that $IS=S^{-1}I, I^2=1$. Imagine that the attracting
set $\AA$ differs from its time reversal image $I\AA=\RR$ and
that CH holds.

The tangent space at a generic point $z$ is supposed to be
smoothly decomposed as $T_u(z)\oplus T_s(z)\oplus T_m(z)$. If
$z\in\AA$ or $z\in \RR$ then $T_s(z),T_u(z)$ coincide with the
tangent, at $z$, to the stable manifold of $S$ on $\AA$ or $\RR$
respectively; furthermore for each ball $U_\d(x)\subset\AA$, of
radius $\d$, consider the manifolds $W_i(x)\cap U_\d(x), i=u,s$,
and assume that they can be continued into smooth manifolds
$W_+,W_-$ everywhere tangent $T_s\oplus T_m$ and $T_u\oplus T_m$
and which intersect $\RR$ in a single point $\wt x= Px$ if $\d$
is small enough: thus defining $P$ as a map between $\AA,\RR$.

Finally, as the labels s,u suggest, the vectors in $T_s,T_u$
uniformly contract exponentially as time tends to $+\infty$ or
$-\infty$ respectively, while vectors in $T_m$ contract
exponentially as $t\to\pm\infty$ (\ie in both directions, being
´squeezed´ on $\AA$ and $\RR$).

The case of a flow $S_t$ can be described similarly by imagining
that $T_m$ contains also the neutral direction $\frac{d}{dt}
S_t\uu$ and contracts transversally to it.  In this context Axiom
C adds to the Axiom B, \cite{Sm967}, the assumption of the
existence of a repeller $\RR$ intersected by the manifolds
emerging from $\AA$.

The latter property permits to establish the map $P$, thus
allowing to define the composition $\wt I=PI$, acting as a time
reversal on $\AA$ and $\RR$, because the invariance of the
manifolds implies that on $\AA\cup\RR$ it is $P S_t=S_tP$: so
that $\wt I S_t=S_{-t}\wt I$ on $\AA\cup\RR$ (note that $\wt I$
is not defined outside $\AA\cup\RR$). See \cite{BG997} for more
details.
\*

\0{\small\bf Acknowledgements:} \rm I am indebted to M. Cencini
for pointing out an error on the first version of the paper, and
to a referee for suggesting improvements.
\*

\0{\bf Erratum:} This is version 3: the former Appendix C has
been suppressed (together with the sentence referring to it
following Eq.\equ{a.7}) because of errors in the attempt to
understand heuristically the apparent approximate symmetry of the
Lyapunov exponents. The former Appendix D is now Appendix C.

\pagina

\bibliographystyle{apalike}

\end{document}